\documentclass[a4paper]{article}

\usepackage{geometry}

\usepackage[utf8]{inputenc}
\usepackage[T1]{fontenc}
\usepackage[english]{babel}
\usepackage[table]{xcolor}
\usepackage{url}

\usepackage{amsmath}
\usepackage{amssymb}
\usepackage{amscd}
\usepackage{amsfonts}
\usepackage{euscript}
\usepackage{longtable}
\usepackage{graphicx}
\usepackage{color}
\usepackage{shadow}
\usepackage{rotating}

\title{\bf Spectrum and separability of mixed 2-qubit X-states}
\author{A. Khvedelidze \textsuperscript{a,b,c,d} and A. Torosyan \textsuperscript{a} \bigskip\\
{\footnotesize{${}^a$}A.Razmadze Mathematical Institute, Iv.Javakhishvili Tbilisi State University, Tbilisi, Georgia}\\
{\footnotesize{${}^b$}Institute of Quantum Physics and Engineering  Technologies, Georgian Technical University, Tbilisi, Georgia} \\
{\footnotesize{${}^c$}Laboratory of Informational Technologies, Joint Institute for Nuclear Research, Dubna, Russia} \\
{\footnotesize{${}^d$} National Research Nuclear University,
MEPhI , Moscow, Russia}
}
\date{\empty}
\begin{document}

\maketitle

\begin{abstract}

The separable mixed   2-qubit $X-$states   are classified in accordance with degeneracies  
in the spectrum of density matrices. It is shown that there are  four classes of separable 
$X\--$states,  among them:  one 4D family, a pair of  2D family  and a single, 
zero-dimensional maximally mixed  state.  

\end{abstract} 

\section*{Introduction}

Consider  the space $\mathfrak{P}_X$ of  $4\times 4$  Hermitian matrices of  the   form:
\begin{equation}
\label{eq:Xmatrix7}
\varrho_{X}:=
\left(
\begin{array}{cccc}
\varrho_{11}& 0 &0& \varrho_{14}\\
0&\varrho_{22} &\varrho_{23}& 0\\
0&\varrho_{32} &\varrho_{33}& 0\\
\varrho_{41}& 0 &0& \varrho_{44}
\end{array}
\right)\,.
\end{equation}
Due to the Hermicity,  the diagonal entries in (\ref{eq:Xmatrix7})
are real numbers,  while  elements of the minor diagonal are pairwise complex conjugate numbers,  $\varrho_{14}=\overline{\varrho}_{14}$ and $\varrho_{23}=\overline{\varrho}_{32}\,.$
Supposing that the matrix $\varrho_{X}$  is semi-positive definite,
\begin{equation}
\varrho_X \geq 0\,,
\end{equation}
and has a unit trace,
\begin{equation}
\mbox{tr}\varrho_X= 1\,,
\end{equation}
the $\varrho_{X}$ can be regarded  as the density matrix of a 4-level quantum system. 
Since non-zero elements in  (\ref{eq:Xmatrix7})  are distributed in a shape similar to  the Latin letter ``X'',  the corresponding quantum states are named as $X-$states.

The 7-dimensional space $\mathfrak{P}_X\,$ represents  a subspace of the 15-dimensional state space  $\mathfrak{P}\,$ of a generic 4-level quantum system, $\mathfrak{P}_X\,\subset\mathfrak{P}\,.$  Since the introduction of $X-$states \cite{YuEberly2007}, various 
subfamilies of $\mathfrak{P}_X\,$  have been attracting  a special attention. There are at least two reasons for that interest.   First of all,  it was found that  microscopic systems,  being  in certain  $X- $states,   show a  highly non-trivial  quantum  behaviour. \footnote{The well-known entangled states,  such as 
Bell states \cite{NielsenChuang}, Werner states \cite{Werner},
isotropic states \cite{HorodeckiHorodecki99}  and maximally entangled mixed states
\cite{IshizakaHiroshima2000,VerstraeteAudenaertBieMoor2001},  are particular subsets of $X-$states.  For further references on $X-$states cf.  \cite{QuesQasJames},  \cite{MendoncaMarchiolliGaletti2014}.}
Secondly, due to the simple algebraic structure  of  $X-$states, many computational  difficulties, common for generic states,  can be resolved dealing  with this special 
subclass of states. \footnote{Such simplifications take place owing  to a discrete symmetry $X-$states possess.   Namely, it can be easily verified that 
every $X-$state (\ref{eq:Xmatrix7})  is  equivalent  to a block-diagonal matrix
\begin{equation}\label{eq:XmatrixBlockDiag}
\varrho_{X} =
P_{\pi}\left(
\begin{array}{cccc}
\varrho_{11}& \varrho_{14} & 0& 0\\
\varrho_{41}& \varrho_{44}&0& 0\\
0& 0&\varrho_{33}&\varrho_{32} \\
0 & 0 &\varrho_{23}& \varrho_{22}\,.
\end{array}
\right)P_{\pi}\,, \quad \mbox{with } \quad 
P_{\pi}= \left[\begin{matrix}
1 & 0 & 0 & 0\\
0 & 0 & 0 & 1\\
0 & 0 & 1 & 0\\
0 & 1 & 0 & 0
\end{matrix}
\right].
\end{equation}}

The aforementioned simplification turned out to be very  important in describing such a complicated phenomenon as the entanglement in composite quantum systems.   Particularly,  it is well-known 
that the famous entanglement  measure - concurrence - can be reduced to a simple  analytical expression for $X-$states.
In the present note we will move towards a detailed  entanglement classification of the mixed 2-qubit $X-$states.
Namely,  the  parametrization of  separable 
mixed $X\--$states  of    two qubits with an  arbitrary spectrum of the density matrix 
will be described.
Our analysis  in the subsequent Sections  includes the following steps:
\begin{enumerate}
\item  Two unitary groups,  both acting adjointly on
the 7-dimensional space of 2-qubit $X-$states, will be introduced;
\begin{enumerate}
\item The first one is the so-called  \textit{``global group''}, $G_X\in  SU(4) \,, $  defined as  the invariance group of the subspace $\mathfrak{P}_X\,,$ 
\[
G_X \varrho_X G_X^\dag \in  \mathfrak{P}_X  \quad \forall \quad     \varrho_X   \in  \mathfrak{P}_X  \,.
\]
\item The second one is the subgroup of  
$G_X\,, $  the so-called \textit{``local group''}, $LG_{{}_X } \in G_X.  $   Its   elements have a tensor product form  corresponding to the decomposition of the state space 
$\mathfrak{P}_X$  into  two qubits subspaces,  $LG_{{}_X} \in SU(2)\times SU(2).$
\end{enumerate}

\item The ``global orbits'',   $\mathcal{O}_{\varrho}$\,, of the group $G_X\,$   will be identified and classified into families/types  according to the degeneracies in  the  spectrum  of  density matrices.

\item Considering  the equivalence classes  induced by the local group $LG_{{}_X} \,$ action on  $\mathcal{O}_{\varrho}\,,$ one can divide the latter into different  subfamilies according to their  entanglement characteristics. Having in mind this ranging,   the separable  density  $X$-matrices will 
be categorized within  the global orbits classification.  
\end{enumerate}

\section{Global and local invariance groups  of $X-$states }

In order to prove  the  properties of  2-qubit  $X-$states announced above, 
let us start with  few definitions.

\noindent{ $\bullet$\bf\, Invariance subalgebra of  $X-$states $\bullet$ }The basis for the $  \mathfrak{su(4)}$ algebra is constructed as follows: 
let $\sigma_\mu = \left( \sigma_0,  \boldsymbol{\sigma}\right)$s denote the set of $2\times 2$
matrices, where $\sigma_0 = I$ is a unit matrix and
$\boldsymbol{\sigma}:=(\sigma_x, \sigma_y, \sigma_z)\,$ are three
Pauli matrices
\begin{equation*}\label{eq:PauliMatices}
    \sigma_x= \left(\begin{array}{cc}
                0 & 1\\
                1 & 0
              \end{array}\right)\,,
              \qquad
\sigma_y= \left(\begin{array}{cc}
                0 & -\imath\\
                \imath & 0
              \end{array}\right)\,,
              \qquad
\sigma_z= \left(\begin{array}{cc}
                1 & 0\\
                0 & -1
              \end{array}\right)\,.
              \qquad
\end{equation*}
The set of  all possible tensor products of  two copies  of  matrices $\sigma_{\mu}\,$,
\[
\sigma_{\mu\nu}:=\sigma_\mu\otimes \sigma_\nu\,, \qquad
\mu, \nu = 0, x,y,z\,,
\]
 forms the basis of the algebra  $\mathfrak{su}(4)\,.$  For our aims it is useful  to write the 
 latter  as the direct sum,
\(
\mathfrak{su(4)}=\mathfrak{l}\oplus\mathfrak{p}\,,
\)
where  the 6-dimensional vector space $\mathfrak{l}$ is composed as 
\begin{equation}\label{eq;lspan}
\mathfrak{l}= \mbox{span}\,\frac{i}{{2}}\{\sigma_{x0},
\sigma_{y0}, \sigma_{z0}, \sigma_{0x}, \sigma_{0y}, \sigma_{0z}\}\,,
\end{equation}
while the 9-dimensional space $\mathfrak{p}$ is 
\footnote{Since the commutators between elements of two subspaces 
$\mathfrak{l}$ and $\mathfrak{p}$  are such that 
\[
[\mathfrak{l}, \mathfrak{l}] \subset \mathfrak{l}\,,
\qquad
[\mathfrak{p}, \mathfrak{l}] \subset \mathfrak{p}\,,
\qquad
 [\mathfrak{p}, \mathfrak{p}] \subset
\mathfrak{l}\,, 
\]
the direct sum $\mathfrak{l}\oplus\mathfrak{p}$ is nothing else than the Cartan decomposition of $\mathfrak{su}(4)$. 
}

\begin{equation}\label{eq;pspan}
\mathfrak{p}= \mbox{span}\, \frac{i}{{2}}\{\sigma_{xx},
\sigma_{xy}, \sigma_{xz}, \sigma_{yx}, \sigma_{yy}, \sigma_{yz},
\sigma_{zx}, \sigma_{zy}, \sigma_{zz}\}\,.
\end{equation}
From now,  to denote the matrices in (\ref{eq;lspan})
and (\ref{eq;pspan}), the notations $\lambda_k\,, $ where $k$ runs from  1 to 15,  will be used
\begin{equation}\label{eq;lambda}
\mathfrak{l}= \mbox{span}\,\{\lambda_1, \lambda_2, \dots,
\lambda_6\}\,, \qquad \mathfrak{p}= \mbox{span}\,\{\lambda_7,
\lambda_8, \dots, \lambda_{15}\}\,.
\end{equation}

$X-$states  (\ref{eq:Xmatrix7})  expand over  the 
subset 
$\alpha_X=\{ \lambda_{15}, \lambda_{10}, \lambda_6,  -\lambda_{11}, \lambda_8, \lambda_3, \lambda_7 \}$
of the  introduced 
$\mathfrak{su(4)}$ basis:
\begin{equation}
\label{eq:Xmatrexp}
\varrho_X= \frac{1}{4}\left(I +2 i\sum_{\lambda_k \in \alpha_X} h_k\lambda_k\right)\,.
\end{equation}
The  real coefficients $h_k$ in (\ref{eq:Xmatrexp}) are given by the linear combinations of the density matrix elements: 
\begin{eqnarray}
 &&h_3    =- \varrho_{11}- \varrho_{22}+ \varrho_{33}+\varrho_{44}\,,  
 \qquad 
 h_6     = -\varrho_{11} +\varrho_{22}-\varrho_{33}+\varrho_{44}\,,\\  
 && h_7    =- \varrho_{14}-\varrho_{23}-\varrho_{32}-\varrho_{41}\,,  
  \qquad 
 h_{11}= - \varrho_{14}+\varrho_{23}+\varrho_{32} - \varrho_{41}\,,\\
 && h_8    = i(-\varrho_{14}+\varrho_{23}-\varrho_{32}+\varrho_{41})  \,,
\quad 
h_{10}= i(-\varrho_{14} -\varrho_{23}+\varrho_{32}+\varrho_{41} )\,,\\
 &&  h_{15}= - \varrho_{11}+\varrho_{22}+\varrho_{33} -\varrho_{44} \,.
\end{eqnarray}
The subset $\alpha_X$ possesses the following properties: 
\begin{itemize}
\item[i.] The subset is closed under the matrix commutator operation, i.e.,  its elements span the  subalgebra of   
$\mathfrak{su(4)}$;
\item[ii.] From the commutators collected in the Table 1.  it follows that the element $\lambda_{15}$  commutes with all other elements of  $\alpha_X$;
\item[iii.] The remaining six   elements, $\{\lambda_3, \lambda_6, \lambda_7,
\lambda_8, \lambda_{10}, \lambda_{11}\}\,$, span the $\mathfrak{su(2)}\oplus\mathfrak{su(2)}$.
\end{itemize}
To check the  last property, one can  construct the following  linear combinations:
\begin{eqnarray}
S_z&=& i(\lambda_3+\lambda_6)\,, 
\qquad
 S_{\pm} = \pm  (\lambda_8+\lambda_{10})+ i(\lambda_7-\lambda_{11})\,,\\
T_z&= &i(\lambda_3-\lambda_6)\,,   \qquad 
T_{\pm} = \mp (\lambda_8-\lambda_{10})+ i(\lambda_7+\lambda_{11})\,,
\end{eqnarray}
and verify that their commutator relations read 
\begin{eqnarray}
&&[S_z,  S_\pm]=\pm 2  S_\pm\,,  \qquad [S_+,  S_-]= 4 S_z\,, \\
&&[T_z,  T_\pm]=\pm 2 T_\pm\,,  \qquad [T_+,  T_-]= 4 T_z\,.
\end{eqnarray}
Thus, two sets of elements 
\begin{eqnarray}
&&\boldsymbol{S}=\{  \frac{1}{2}(S_+ + S_-),  \  \frac{i}{2}(S_+-S_- ),\   S_z   \}\,,    \\
&&\boldsymbol{T}=\{  \frac{1}{2}(T_+ + T_-), \   \frac{i}{2}(T_+-T_- ),\  T_z   \}\,
\end{eqnarray}
generate two copies of $\mathfrak{su(2)}$ algebra. 
\footnote{In  terminology of \cite{Rau2000} such  operators describe ``pseudospins'' for two-spin system.}
Gathering all together,
we conclude that the set $\alpha_X$ generates the  subalgebra 
$\mathfrak{g}_X:=\mathfrak{su(2)}\oplus\mathfrak{su(2)} \oplus\mathfrak{u(1)} \in \mathfrak{su(4)}.$
\footnote{For further information on a diverse algebraic structure of $X$-states 
see \cite{Rau2009}.}

\noindent{ $\bullet$\bf\, Global unitary group of   $X-$states $\bullet$ } 
Exponentiation   of  the algebra $ \mathfrak{g}_X$ results in the 7-parametric  subgroup of $SU(4)$\,,
\[
G_X :=\exp (\mathfrak{g}_X)\in SU(4)\,,
\]
whose action  preserves the  $X-$states space  $\mathfrak{P}_X$, i.e., 
\(
G_X \varrho_X G_X^\dag \in  \mathfrak{P}_X\,.
\)
Using  the expansion $\mathfrak{g}_X = \sum_i \omega_i\lambda_i\,  $ over the 7-tuple $\lambda_i \in \alpha_X\,$   and   the formulae (\ref{eq:ChB1})-(\ref{eq:ChB2}) from  the  Section {\bf \ref{sec:Supmat}. Supplementary material}, 
one can verify that   the group   $G_X$  has the following representation:
\begin{equation}\label{eq:G_XRepr}
G_X=P_{\pi}\left(
\begin{array}{c|c}
{e^{-i {\omega_{15}}}SU(2)}& 0  \\
\hline
0 &{e^{i {\omega_{15}}}SU(2)^\prime}\\
\end{array}
\right)P_{\pi}\,,
\end{equation}
where the two copies of SU(2) are parametrized as follows:
\begin{eqnarray*}
SU(2)&=& \exp{\left[i\left(\omega_4+\omega_7\right)\sigma_1 +i\left(\omega_2+\omega_5\right)\sigma_2 +
i\left(\omega_3+\omega_6\right)\sigma_3  \right]}\,,\\
SU(2)^\prime&=& \exp{\left[i\left(-\omega_4+\omega_7\right)\sigma_1 +i\left(-\omega_2+\omega_5\right)\sigma_2 +
i\left(\omega_3 -\omega_6\right)\sigma_3 \right] }\,.
\end{eqnarray*}

\noindent{ $\bullet$\bf\,  Local subgroup of  $ G_X$  $\bullet$ }
Suppose now that our  4-level system is composed of  2-level subsystems, i.e.,   two qubits. 
In this case 
the Hilbert space $\mathcal{H}$  is given by  the tensor  product of  2-dimensional Hilbert spaces, 
\(
\mathcal{H}= \mathcal{H}_1\otimes \mathcal{H}_2\,,
\)
and one can consider  the tensor product of operators acting independently on  the subspaces  of individual qubits,  $\mathcal{H}_1$ and $\mathcal{H}_2\,.$
Particularly, having in mind  an intuitive idea of mutual independence of isolated qubits,  we define  
the \textit{``local unitary group''}, $LG_{{}_X } \,$, as the subgroup of  global invariance group of $X-$ states, $G_{{}_X} \,,$    such that each of its elements $g  \in LG_{{}_X }$ has  the tensor product form,   $g = g_1\times g_2\,,$  with  $g_1, g_2 \in SU(2)\,. $
From the expression  (\ref{eq:G_XRepr}) it follows that  the local unitary group can be written as:
\begin{equation}
LG_{{}_X }=P_{\pi}\exp(\imath\ \frac{\varphi _1}{2}\sigma_3) \times \exp(\imath\frac{\varphi _2}{2} \sigma_3) P_{\pi} \,.
\end{equation}

\section{Global   $G_{{}_X}-$orbits }

Now  it will be shown that every $X-$states density matrix can be diagonalized
using  some subgroup of global $G_{{}_X} $ group.  
Therefore,  the adjoint $G_{{}_X}\-- $orbits structure is completely  determined by  the coset
$G_{{}_X} / H_\varrho$,  where   $H_\varrho$  stands for the isotropy group of a density matrix $\varrho$. 
This isotropy  group, in turn,  depends on  the degeneracies occurring  in the spectrum of density matrices. Thus, the 
latter determines   all possible types of $G_{{}_X}\-- $orbits and the corresponding classification can be carried as follows.

\subsection{Dimensionality of the tangent space of $G_{{}_X}\-- $orbits }

Consider  the adjoint action of the global unitary group $G_{{}_X}$  on the  7-dimensional space 
$\mathfrak{P}_X\,$
and introduce the following vectors at each point $\varrho \in \mathfrak{P}_X$: 
\begin{equation}\label{eq:tangVec}
t_k=\frac{\partial}{\partial v_k} \left(g(\boldsymbol{v})\varrho_X g^{\dagger}(\boldsymbol{v})\right) \bigg|_{v_k=0} =  [\lambda_k,\varrho_X]\,,  \quad  k= 3, 6, 7, 8, {10},  {11},  {15}\,.
\end{equation}
In the equation (\ref{eq:tangVec})  the group elements $g(\boldsymbol{v})\in G_{{}_X}$ are parametrized by 7-tuple $\boldsymbol{v}=\{v_3, v_6, v_7, v_8, v_{10}, v_{11},  v_{15} \}$:
\begin{equation}
g(\boldsymbol{v}) =\exp\left( \sum_{\lambda_k\in \alpha_X} v_k \lambda_k\right)\,.
\end{equation}
These vectors  belong to a tangent space of $G_{{}_X}$-orbits. 
The dimensionality of this  tangent space is given by the rank of the  $7 \times 7$ Gram matrix
\begin{equation}
G = \|G_{kl}\| = \frac{1}{2} \|Tr(t_k t_l)\|\,.
\end{equation}
Straightforward evaluation of the spectrum $\sigma(G)$ of the Gram matrix $G$ shows that it comprises:  two eigenvalues of multiplicity 2 and three identically vanishing eigenvalues, 
\begin{equation}
\sigma(G)=\{\mu_1, \mu_1, \mu_2,\mu_2,0,0,0\}\,,
\end{equation}
where the double multiplicity eigenvalues read: 
\begin{eqnarray}\label{eq:dmeigen1}
\mu_1&=& \left(h_3+h_6\right){}^2+\left(h_8+h_{10}\right){}^2+\left(h_7+h_{11}\right){}^2\,,\\
\mu_2&= &\left(h_3-h_6\right){}^2+\left(h_8-h_{10}\right){}^2+\left(h_7-h_{11}\right){}^2\,. 
\label{eq:dmeigen2}
\end{eqnarray}
The formulae  (\ref{eq:dmeigen1}) and (\ref{eq:dmeigen2})  ensure  that there exist 4 types of $G_{{}_X}$-orbits:
\begin{itemize}
\item{\bf dim $\mathcal{O}$ =4}, the generic orbits;    
\item {\bf dim $\mathcal{O}$ =2}, the degenerate orbits  defined  by the equations:
\begin{equation}
h_6=h_3\,, \  h_{10}=h_8\,, \  \  h_{11}=h_7\,;
\end{equation}
\item {\bf dim $\mathcal{O}$ =2}, the degenerate orbits 
defined by the equations:
\begin{equation}
h_6=-h_3\,, \  h_{10}=-h_8\,, \  \  h_{11}=-h_7\,;
\end{equation}
 \item {\bf dim $\mathcal{O}$ =0}, the single orbit  $\varrho_X= \frac{1}{4 }I \,$ -  the  maximally mixed state. 

\end{itemize}
In terms of the eigenvalues of density matrices,  the 4D orbits  are consistent with 
a generic spectrum, i.e., matrices with  4 different eigenvalues,  while 2D orbits  are generated  by 
  $X$-matrices with double multiplicity of the following types: 
\begin{equation}\label{eq:XmatrixDegenerat}
P_{\pi}\left(
\begin{array}{cccc}
\varrho_{11}& \varrho_{14} &0& 0\\
\varrho_{41}& \varrho_{44}&0& 0\\
0& 0&\varrho_{22}& 0\\
0 & 0 &0& \varrho_{22}\,.  
\end{array}
\right)P_{\pi}
 \
 \mbox{and}
\
P_{\pi}\left(
\begin{array}{cccc}
\varrho_{11}& 0&0& 0\\
0& \varrho_{11}&0& 0\\
0& 0&\varrho_{33}&\varrho_{32} \\
0 & 0 &\varrho_{23}& \varrho_{22}\,.
\end{array}
\right)P_{\pi}\,.
\end{equation}

\subsection{$G_{{}_X}\-- $orbits parametrization }

Here  a detailed  representation for each type of $G_{{}_X}\-- $orbits  
will be given, starting from  the orbit of the highest  dimensionality.

\subsubsection{Generic orbits, $\mbox{dim}\left( {\mathcal{O}}\right)=4$}

Let us assume that the spectrum of $\varrho_X$  is  a generic one, i.e.,  all eigenvalues  
$\sigma(\varrho):=\{ r_1,  r_2, r_3,  r_4 \}$ are different positive  real numbers.  
Furthermore,  in the block-diagonal representation (\ref{eq:XmatrixBlockDiag}) of the density matrix  $\varrho_X$,  the 
$\{r_1,  r_2\}$ denote the eigenvalues of the upper block and $\{r_3,  r_4\}$ 
are eigenvalues of the lower block. 

\begin{figure}[h]
\centering
\includegraphics[scale=0.4]{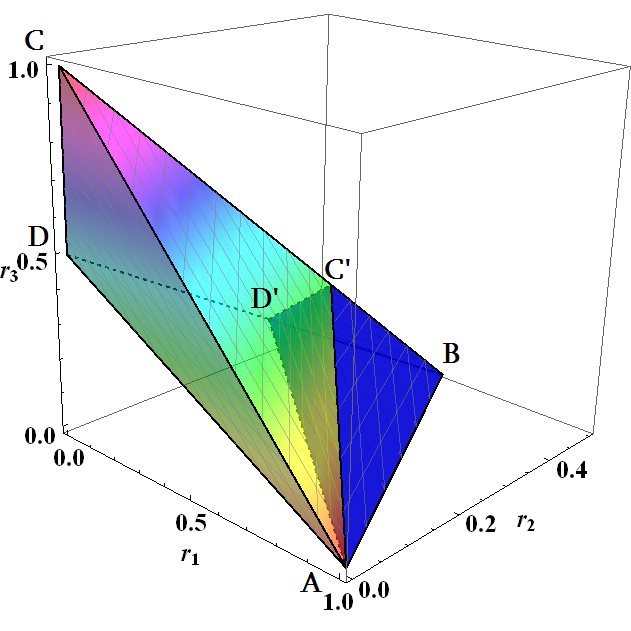}
\caption{The tetrahedron $ABCD$ as the image of the  partially ordered simplex  $\underline{\Delta}_3$,  while the tetrahedron $ABC'D'$ inside  it  corresponds to a 3D simplex 
 with the following complete order of eigenvalues: 
 $ \{  \  \sum_{i=1}^4 r_i =1 \,, \quad   1 \geq r_1 \geq r_2 \geq r_3 \geq r_4 \geq  0\ \} \,$.}{ \label{Fig:PartOrderedSimplex}}
\end{figure}

The $4\times 4 $ density matrix $\varrho_X$  can be diagonalized in a blockwise way, 
\begin{equation} \label{eq:ddiagXmatr}
\varrho_X= W \left(
\begin{array}{c|c}
$\mbox{diag}$(r_1, r_2) & 0 
\\\hline
 0 &$\mbox{diag}$(r_3, r_4)\\
\end{array}
\right) W^\dag\,, 
\end{equation}
using  a special unitary matrix 
\begin{equation} \label{eq:UnitDiag}
W = P_{\pi}\left(
\begin{array}{c|c}
e^{i\omega}{\displaystyle{U}}& {}^{\mbox{\Large  0 }} 
\\\hline
 {}_{\mbox{\Large  0 }}&e^{-i\omega}{ \displaystyle{V}}\\
\end{array}
\right) P_{\pi}, 
\end{equation}
where $U$ and $V$  are $2\times 2$ special unitary matrices diagonalizing the upper and lower  sub-blocks in (\ref{eq:XmatrixBlockDiag}). 
Since a generic spectrum has been assumed,  matrices $U$ and $V$ belong to 
the  coset,  $SU(2)/ U(1)\times S_2$, where the  group $S_2$ interchanges  eigenvalues 
inside the pairs $\{ r_1,  r_2\}$ and $\{r_3,  r_4 \}$. In order to have uniqueness in (\ref{eq:ddiagXmatr}),   one can fix a certain order in the spectrum 
$\sigma(\varrho_X).$  Namely,  we assume that  elements of the spectrum form a 
partially ordered simplex,  $\underline{\Delta}_3$, i.e.,  
\begin{equation}\label{eq:semi-orderedSim}
\underline{\Delta}_3: \quad \sum_{i=1}^4 r_i =1 \,, \qquad 0 \leq r_2  \leq  r_1 \leq 1 ,
 \quad 
 0 \leq r_4  \leq  r_3 \leq 1 \,,
\end{equation}
depicted in the {\sc  Figure} \ref{Fig:PartOrderedSimplex}. 
\footnote{Note that the case  of general position considered  here consists of points inside the $\underline{\Delta}_3$ and  satisfies  the inequalities $  r_2  < r_1 $ and $r_4  <r_3 \,.$
}

 Comparing expression  (\ref{eq:UnitDiag})  with   (\ref{eq:G_XRepr}),  we convinced  that
the diagonalizing  matrix  is an element of the global group $W \in G_{{}_X}\,$
with $2\times 2 $ special unitary matrices $U$ and $V$ from the coset 
$SU(2)/ U(1)$ parametrized by angles $\phi_1,\phi_2 \in [0, \pi]\,, \psi_1,\psi_2 \in [0, 2\pi]$: \begin{equation}\label{eq:UV}
 U =e^{i \displaystyle{\frac{\psi_1}{2}\sigma_3}}e^{i \displaystyle{\frac{\phi_1}{2}\sigma_2}}\,,
 \qquad
 V=e^{i \displaystyle{\frac{\psi_2}{2}\sigma_3}}e^{i \displaystyle{\frac{\phi_2}{2}\sigma_2}}\,.
 \end{equation}
 The 3-dimensional  isotropy group $H_{\mbox{Generic}}$ of generic orbits is  
 \begin{equation}
 H_{\mbox{Generic}}=P_{\pi}\left(
\begin{array}{c|c}
e^{i\omega}\exp{\displaystyle{\frac{\gamma_1}{2}\sigma_3}}& {}^{\mbox{\Large  0 }} 
\\\hline
 {}_{\mbox{\Large  0 }}&e^{-i\omega}\exp{ \displaystyle{\frac{\gamma_2}{2}\sigma_3}}\\
\end{array}
\right) P_{\pi}\,. 
 \end{equation}
 This is in accordance with the  maximal dimension of  the  $G_{{}_X}$-orbits:
 $$
\mbox{dim}\left( {\mathcal{O}}\right)_{\mbox{Generic}}= \mbox{dim}\left(G_{{}_X }\right) -
\mbox{dim} H_{\mbox{Generic}}
=7-3=4 \,.
$$
Summarising,    the adjoint action of the  global group $ G_{{}_X}$  determines the generic orbits,  
which are locally given by product of  2-spheres,
$S_2\times S_2 $.

 \subsubsection{Degenerate  orbits, $\mbox{dim}\left( {\mathcal{O}}\right)=2$}
 
According to the representation (\ref{eq:XmatrixDegenerat}), two types of  2D  degenerate $G_{{}_X}$-orbits are generated by the matrices  with degenerate $2\times 2$ sub-blocks,   either  upper or  lower blocks. 
In the first case the  isotropy group $H_{\mbox{\small Degenerate }}$  reads 
 \begin{equation}
 H_{\mbox{\small Degenerate}}=P_{\pi}\left(
\begin{array}{c|c}
e^{i\omega}SU(2)& {}^{\mbox{\Large  0 }} 
\\\hline
 {}_{\mbox{\Large  0 }}&e^{-i\omega}\exp{ \displaystyle{\frac{\gamma_2}{2}\sigma_3}}\\
\end{array}
\right) P_{\pi}\,, 
 \end{equation}
 while for the second case $H_{\mbox{\small Degenerate }}$  is 
 \begin{equation}
 H^\prime_{\mbox{ \small Degenerate}}=P_{\pi}\left(
\begin{array}{c|c}
e^{i\omega}\exp{\displaystyle{\frac{\gamma_1}{2}\sigma_3}}& {}^{\mbox{\Large  0 }} 
\\\hline
 {}_{\mbox{\Large  0 }}&e^{-i\omega}SU(2)^\prime\\
\end{array}
\right) P_{\pi}\,. 
 \end{equation}
 In both cases,  $
\mbox{dim} H_{\mbox{\small Degenerate}}=\mbox{dim} H^\prime_{\mbox{\small Degenerate}}= 5 \,
$
 and the dimension of  these degenerate $G_{{}_X}$-orbits is 
 $$
\mbox{dim}\left( {\mathcal{O}}\right)_{\mbox{\small Degenerate}}= \mbox{dim}\left(G_{{}_X }\right) -
\mbox{dim} H_{\mbox{\small Degenerate}}
=7-5=2 \,.
$$
 \subsubsection{Degenerate  orbit, $\mbox{dim}\left( {\mathcal{O}}\right)=0$}
 
Finally, there is one point  in the state space $\mathfrak{P}_X\,$, whose  isotropy group coincides with the invariance group $G_{{}_X}\,.$ This point corresponds to the maximally mixed state,
$\varrho_X= \frac{1}{4} I\,.$

\section{The separable  states}

Now we are in position to  prove that every type of $G_{{}_X}$-orbits
includes  the separable states.
\footnote{
The density matrix $\varrho\,$ describing the mixed state of a composed  system 
$\mathcal{H}= \mathcal{H}_1\otimes \mathcal{H}_2, $ is \textit{separable} if it allows the convex 
 decomposition:
 \begin{equation}
\varrho =\sum_{k} \omega_k \varrho_1^k\otimes\varrho_2^k\,, \qquad \sum_{k}\omega_k =1,\quad \omega_k >0\,,
\end{equation}
where $\varrho^k_1 $ and $\varrho^k_2$ represent the density matrices  acting  
on  the multipliers  $\mathcal{H}_1$ and $\mathcal{H}_2$ correspondingly. Otherwise, it is \textit{entangled} \cite{Werner} .
}

\begin{figure}[h]
\centering
\includegraphics[scale=0.4]{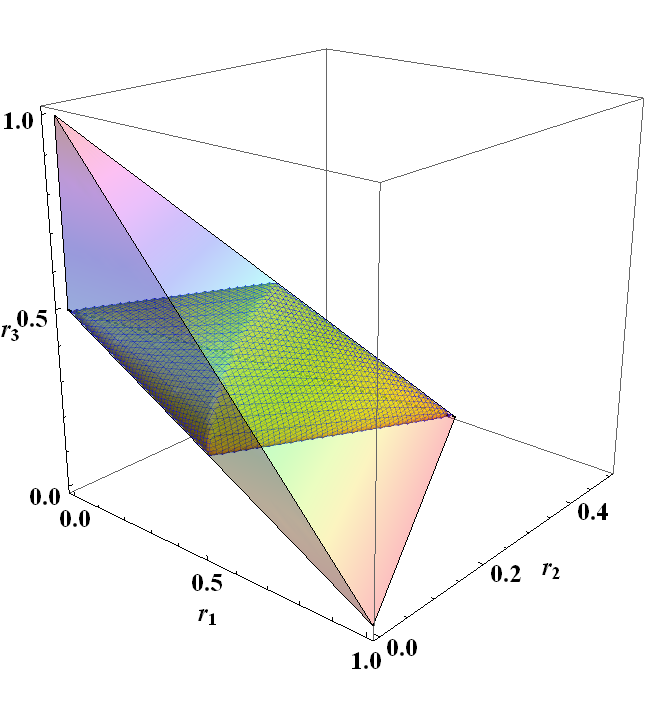}
\caption{The absolute separable states inside the X-states tetrahedron.}{ \label{Fig:AbsSepX}}
\end{figure}

\subsection{Separable states on the generic $G_{{}_X}$-orbits}

 The separability of states as a function of density matrices spectrum $\sigma(\varrho_X)$, can be analysed using the  representation   (\ref{eq:ddiagXmatr}) for the generic $G_{{}_X}$-orbits.

According to the Peres-Horodecki criterion \cite{Peres1996}, which is a necessary and sufficient condition for separability of $2\times 2$ and $2\times 3$ dimensional systems, a state $\varrho$ is separable if its partial transposition, i.e.,   $\varrho^{T_2} = I\otimes T \varrho\,,$  is
 semi-positive as well.\footnote{Here we consider the partial transposition  with respect  to the  ordinary 
transposition operation    $T$  in the second subsystem; 
similarly, one can use the alternative action,  $\varrho^{T_1} = T\otimes I \varrho.$}
Straightforward computation with $\varrho_X\, $  in the form (\ref{eq:ddiagXmatr})  shows that  the semi-positivity of the partially transposed matrix $\varrho_X^{T_2}\, $  requires   fulfilment of the following inequalities:
 \begin{eqnarray}\label{eq:1posTREig}
&&(r_1-r_2)^2\cos^2\phi_1+(r_3-r_4)^2 \sin^2\phi_2 \leq (r_1+r_2)^2,
\\
&&
(r_3-r_4)^2\cos^2\phi_2+(r_1-r_2)^2 \sin^2\phi_1 \leq (r_3+r_4)^2.
\label{eq:2posTREig}
\end{eqnarray}

Note that the inequalities   (\ref{eq:1posTREig}) and (\ref{eq:2posTREig}) do not constraint  
 two angles $\psi_1$ and $\psi_2 $  in (\ref{eq:UV}) that parametrize  the   local group  
 $K = \exp{(i\frac{\psi_1}{2} \sigma_3)}\times \exp{(i \frac{\psi_2}{2} \sigma_3)}. $ It conforms with a general observation that the separability  property is independent from the  local characteristics of the composite system. 
 This local group  is  the factor of  the global group   $G_{{}_X}=K G^\prime_{{}_X} \, $,  and  the corresponding factor in the matrix  $W$  diagonalising $\varrho_X$, is irrelevant  for the separability of $X$- states.

Analysing the inequalities (\ref{eq:1posTREig}) and (\ref{eq:2posTREig}), one can conclude:
\begin{itemize}
\item[i.] There are separable states for any values of eigenvalues from the partially ordered simplex $\underline{\Delta}_3$. In other words, the inequalities (\ref{eq:1posTREig}) and 
(\ref{eq:2posTREig}) determine non-empty domain of definition for  angles 
$\phi_1$ and $\phi_2$ in (\ref{eq:UV}) for every non-degenerate  spectrum $\sigma(\varrho_X)$;
\item[ii.] There is a special family  of the  so-called ``absolutely  separable'' $X$-states,  such that 
the  angles  $\phi_1$ and $\phi_2$ can be arbitrary one. 
The absolutely separable $X$-states are generated by subset  of the partially ordered simplex (\ref{eq:semi-orderedSim}) defined by the inequalities 
 \begin{eqnarray}\label{eq:absSepX}
&& (r_1-r_2)^2  \leq  4r_3r_4\,,\\
 && (r_3-r_4) ^2  \leq 4r_1r_2\,.
 \end{eqnarray}
\end{itemize}
The {\sc Figure }\ref{Fig:AbsSepX}. illustrates location of the subset of the  absolutely separable states 
inside the partially ordered simplex $\underline{\Delta}_3\,.$

\subsection{Separable states on the degenerate  $G_{{}_X}$-orbits}

Testing the degenerate density matrices of the form (\ref{eq:XmatrixDegenerat}) by 
the Peres-Horodecki criterion,  we reveal  the following picture. 
The positivity requirement of partially transposed  density matrix 
with double multiplicity of eigenvalues  gives the  inequalities similar  
to (\ref{eq:1posTREig}) and (\ref{eq:2posTREig}).
However,  owing to the larger isotropy group $H_{\mbox{\small Degenerate }}$ of states,  the new inequalities 
depend solely  on a single  coordinate of  the  coset 
$G_{{}_X}/ H_{\mbox{\small Degenerate}}\,.$
More precisely,  if $r_1=r_2 \,,$ i.e., the degeneracy occurs  in the upper 
sub-block,  then the  angle $\phi_2$ that  parametrizes  the matrix $V$ in (\ref{eq:UV}) 
plays the  role of such a coordinate. 
In this case, the Peres-Horodecki criterion  asserts  that  the degenerate $X-$state is separable iff:
\begin{equation}\label{eq:pos2D}
\cos^2 \phi_2  \leq  \frac{4\zeta}{(1-\zeta)^2} \,,
\end{equation} 
where  $\zeta={r_4}/{r_3} < 1\,$.
This inequality points out the critical value $\zeta_* = 3-2\sqrt{2}\, $, such that 
for $\zeta\leq \zeta_* $ the angle $\phi_2$ is constrained,  while for 
the interval $\zeta_*< \zeta <1 $ the state is separable for an arbitrary angle $\phi_2\,.$
The analogous results for the angle $\phi_1$  (see  the matrix $U$ in (\ref{eq:UV}))
 hold  true if  the lower sub-block in  (\ref{eq:XmatrixDegenerat}) is degenerate, i.e.,   $r_3=r_4\,.$ 
Therefore,  in both classes of  the degenerate  2D  global orbits one can point out 2D  family of separable  degenerate states.  Furthermore,  among them there are  the  ``degenerate absolutely separable'' states, i.e., the  
degenerate global 2D orbits consisting completely  from the  separable states.

\section{Concluding remarks}

The present article is devoted to the discussion of an interplay between local and global characteristics of a pair of qubits in mixed  $X\--$states. 
With this aim, orbits of the global unitary group $G_{{}_X}$ action were described and classified  according to the degeneracies occurring in  the spectrum of density matrices.  Based on this analysis,   
the dependence of $X\--$states  separability  on  the spectrum  has been studied.  
Particularly,  the separable $X\--$states  have been  collected into the following families:
\begin{itemize}
\item The 4-dimensional family of separable states with the spectrum in general position;
\item Two classes of  2-dimensional separable states with the double degeneracy spectrum;
\item The maximally mixed state.
\end{itemize}
Finalizing notes,   it is worth to comment that  according to the aforementioned  classification,  the entangled states being  complementary to the separable  states,  are   partitioned likewise into  three types.  However,  such classification is  not complete.  A further,  more subtle  ranging of  the entangled states located on the given  $G_{{}_X}$-orbit  into subclasses is necessary. The latter subclasses are determined  not by invariants of the global group $G_{{}_X}$, but   are specified by the values of  the $LG_{{}_X}$-invariants.
In the forthcoming publications we are planning to discuss this issue in more detail. Apart from that,  following the approach elaborated in 
\cite{GKP}, \cite{GKPYPh} and \cite{GKPYXstates}, the generalization of the derived results for a  
generic case of 15-dimensional 2-qubit states will be considered.

\section{Supplementary material}
\label{sec:Supmat}

Here we collect a technical material  useful for performing computations 
described in the main text. It includes the basis of the 
Lie algebra $\mathfrak{su}(4)$,   commutators  of its elements  and  block-diagonal  
representation for the subalgebra 
$\alpha_X$.
\bigskip

\noindent{ $\bullet$\bf Basis for  the Lie algebra $\mathfrak{su}(4)$ $\bullet$ }
The anti-Hermitian matrices, 
$$
\{\lambda_1, \lambda_2, \dots, \lambda_6\}= \frac{i}{{2}}\{\sigma_{x0},
\sigma_{y0}, \sigma_{z0}, \sigma_{0x}, \sigma_{0y}, \sigma_{0z}\}\,
$$
and 
$$\{\lambda_7, \lambda_8, \dots, \lambda_{15}\}= \frac{i}{{2}}\{\sigma_{xx},
\sigma_{xy}, \sigma_{xz}, \sigma_{yx}, \sigma_{yy}, \sigma_{yz},
\sigma_{zx}, \sigma_{zy}, \sigma_{zz}\}\,,
$$ read: 

\bigskip
\noindent$
\lambda_1=\frac{i}{2}
\begin{Vmatrix}
        0 & 0 & 1 & 0 \\
        0 & 0 & 0 & 1 \\
        1 & 0 & 0 & 0 \\
        0 & 1 & 0 & 0 \\
 \end{Vmatrix},
\quad
\lambda_2=\frac{i}{2}
 \begin{Vmatrix}
      0 & 0 & -i & 0 \\
      0 & 0 & 0 & -i \\
      i & 0 & 0 & 0 \\
      0 & i & 0 & 0 \\
    \end{Vmatrix},
   \quad
  \lambda_3=\frac{i}{2} 
  \begin{Vmatrix}
      1 & 0 & 0 & 0 \\
      0 & 1 & 0 & 0 \\
      0 & 0 & -1 & 0 \\
      0 & 0 & 0 & -1 \\
     \end{Vmatrix}, 
$
\bigskip
\bigskip

\noindent$
\lambda_4=\frac{i}{2}
\begin{Vmatrix}
       0 & 1 & 0 & 0 \\
       1 & 0 & 0 & 0 \\
       0 & 0 & 0 & 1 \\
       0 & 0 & 1 & 0 \\
      \end{Vmatrix},
   \quad
\lambda_5=\frac{i}{2}
 \begin{Vmatrix}
      0 & -i & 0 & 0 \\
      i & 0 & 0 & 0 \\
      0 & 0 & 0 & -i \\
      0 & 0 & i & 0 \\
     \end{Vmatrix},
     \quad
  \lambda_6=\frac{i}{2} 
      \begin{Vmatrix}
      1 & 0 & 0 & 0 \\
      0 & -1 & 0 & 0 \\
      0 & 0 & 1 & 0 \\
      0 & 0 & 0 & -1 \\
     \end{Vmatrix},
     $
 \bigskip
\bigskip

\noindent$
   \lambda_7=\frac{i}{2} 
     \begin{Vmatrix}
      0 & 0 & 0 & 1 \\
      0 & 0 & 1 & 0 \\
      0 & 1 & 0 & 0 \\
      1 & 0 & 0 & 0 \\
     \end{Vmatrix},
     \quad
  \lambda_8=\frac{i}{2}  
  \begin{Vmatrix}
      0 & 0 & 0 & -i \\
      0 & 0 & i & 0 \\
      0 & -i & 0 & 0 \\
      i & 0 & 0 & 0 \\
     \end{Vmatrix},
     \quad
  \lambda_9=\frac{i}{2}  \begin{Vmatrix}
      0 & 0 & 1 & 0 \\
      0 & 0 & 0 & -1 \\
      1 & 0 & 0 & 0 \\
      0 & -1 & 0 & 0 \\
     \end{Vmatrix}, 
$
\bigskip 
\bigskip

\noindent$
\lambda_{10}=\frac{i}{2}
\begin{Vmatrix}
      0 & 0 & 0 & -i \\
      0 & 0 & -i & 0 \\
      0 & i & 0 & 0 \\
      i & 0 & 0 & 0 \\
     \end{Vmatrix},
\quad
\lambda_{11}=\frac{i}{2}
\begin{Vmatrix}
      0 & 0 & 0 & -1 \\
      0 & 0 & 1 & 0 \\
      0 & 1 & 0 & 0 \\
      -1 & 0 & 0 & 0 \\
     \end{Vmatrix},
     \quad
\lambda_{12}=\frac{i}{2}
 \begin{Vmatrix}
      0 & 0 & -i & 0 \\
      0 & 0 & 0 & i \\
      i & 0 & 0 & 0 \\
      0 & -i & 0 & 0 \\
     \end{Vmatrix},   
$
\bigskip 
\bigskip

\noindent$
\lambda_{13}=\frac{i}{2} \begin{Vmatrix}
      0 & 1 & 0 & 0 \\
      1 & 0 & 0 & 0 \\
      0 & 0 & 0 & -1 \\
      0 & 0 & -1 & 0 \\
     \end{Vmatrix},
\quad
\lambda_{14}=\frac{i}{2}\begin{Vmatrix}
      0 & -i & 0 & 0 \\
      i & 0 & 0 & 0 \\
      0 & 0 & 0 & i \\
      0 & 0 & -i & 0 \\
     \end{Vmatrix},
     \quad
\lambda_{15}=\frac{i}{2}\begin{Vmatrix}
      1 & 0 & 0 & 0 \\
      0 & -1 & 0 & 0 \\
      0 & 0 & -1 & 0 \\
      0 & 0 & 0 & 1 \\
     \end{Vmatrix}. 
$
\bigskip
\bigskip

\begin{sidewaystable}
\centering
{\Large
\begin{tabular}{|>{\columncolor{green}} c|| c |c|>{\columncolor{yellow}} c|c|c|c|c|c|c|c|c|c|c|c
||>{\columncolor{yellow}} c |c}
\hline
\hline
\rowcolor{green}& $\lambda_1 $&$ \lambda_2$ &\cellcolor{yellow}$\lambda_3 $&$\lambda_4$ 
&$\lambda_5$ &\cellcolor{yellow}$\lambda_6 $&\cellcolor{yellow}$\lambda_7$
&\cellcolor{yellow}$\lambda_8$ 
&$\lambda_9$ &\cellcolor{yellow}$\lambda_{10}$ &\cellcolor{yellow}$\lambda_{11}$
&$\lambda_{12}$ &$\lambda_{13}$ &$\lambda_{14} $&\cellcolor{yellow}$\lambda_{15}$ \\ \hline\hline
$\lambda_1 $&  0  & -$\lambda_3 $ & \cellcolor{white} $\lambda_2$ & 0 &0&0&0&0 &0&-$\lambda_{13}$&-$\lambda_{14}$&-$\lambda_{15}$&$\lambda_{10}$&$\lambda_{11}$
&\cellcolor{white}$\lambda_{12}$\\ \hline
$\lambda_2 $ & $\lambda_3$ &0&\cellcolor{white}-$\lambda_1$&\cellcolor{white}0&0&0& $\lambda_{13}$ &$\lambda_{14}$&$\lambda_{15}$&0&0&0&-$\lambda_7$&-$\lambda_8$&\cellcolor{white}$-\lambda_{9}$\\\hline \hline
\rowcolor{yellow}$\lambda_3$&\cellcolor{white}-$\lambda_2$&\cellcolor{white}$\lambda_1$&0
&\cellcolor{white}0&\cellcolor{white}0&0&-$\lambda_{10}$&-$\lambda_{11}$&\cellcolor{white}-$\lambda_{12}$&$\lambda_7$&$\lambda_8$&\cellcolor{white}$\lambda_9$&\cellcolor{white}
0&\cellcolor{white}0&0\\\hline\hline 
$\lambda_4$ & 0 & 0 &\cellcolor{white}0&0&-$\lambda_6$&$\lambda_5$&0&-$\lambda_9$&$\lambda_8$&0&-$\lambda_{12}$&$\lambda_{11}$&0&-$\lambda_{15}$&\cellcolor{white}$\lambda_{14}$
\\\hline
 $\lambda_5$&0 &0&\cellcolor{white}0&$\lambda_6$&0&-$\lambda_4$&$\lambda_9$&0&-$\lambda_7$&$\lambda_{12}$&0&-$\lambda_{10}$&$\lambda_{15}$&0&\cellcolor{white}$\lambda_{13}$\\
\hline 
\hline\rowcolor{yellow}$\lambda_6$&\cellcolor{white} 0 &\cellcolor{white} 0& 0 &\cellcolor{white}-$\lambda_5$&\cellcolor{white}$\lambda_4$&0&-$\lambda_8$&$\lambda_7$
&\cellcolor{white}0&-$\lambda_{11}$&$\lambda_{10}$&\cellcolor{white}0&\cellcolor{white}-$\lambda_{14}$
&\cellcolor{white}$\lambda_{13}$&0\\\hline
\rowcolor{yellow}$\lambda_7$&\cellcolor{white} 0 & \cellcolor{white}-$\lambda_{13}$ & $\lambda_{10}$& \cellcolor{white} 0&\cellcolor{white}-$\lambda_9$ &$\lambda_8$ &0&-$\lambda_6$&\cellcolor{white} $\lambda_5$&-$\lambda_3$&0&\cellcolor{white}0&\cellcolor{white}$\lambda_2$&\cellcolor{white}0&0\\
\hline
\rowcolor{yellow}$\lambda_8$&\cellcolor{white} 0&\cellcolor{white} -$\lambda_{14}$&$\lambda_{11}$ &\cellcolor{white}$\lambda_9$&\cellcolor{white}0&-$\lambda_7$&$\lambda_6$&0&\cellcolor{white}-$\lambda_4$&0&-$\lambda_3$&\cellcolor{white}0&\cellcolor{white}0
&\cellcolor{white}$\lambda_2$& 0\\\hline\hline
$\lambda_9$&\cellcolor{white}0&\cellcolor{white}-$\lambda_{15}$&
\cellcolor{white}$\lambda_{12}$
&-$\lambda_{8}$&$\lambda_7$&0&-$\lambda_5$&$\lambda_4$&0&0&0&-$\lambda_3$&0&0&\cellcolor{white}$\lambda_2$\\\hline\hline
\rowcolor{yellow}$\lambda_{10}$&
\cellcolor{white}$\lambda_{13}$&\cellcolor{white}0&-$\lambda_7$&\cellcolor{white}0&\cellcolor{white}-$\lambda_{12}$&$\lambda_{11}$&$\lambda_3$&0&\cellcolor{white}0&0&-$\lambda_6$
&\cellcolor{white}$\lambda_5$&\cellcolor{white}-$\lambda_1$&\cellcolor{white}0&0\\\hline
\rowcolor{yellow}$\lambda_{11}$ &\cellcolor{white}$\lambda_{14}$&\cellcolor{white}0&-$\lambda_8$&\cellcolor{white}$\lambda_{12}$&\cellcolor{white}0&-$\lambda_{10}$&0&$\lambda_3$&\cellcolor{white}0&$\lambda_6$&0
&\cellcolor{white}-$\lambda_4$&\cellcolor{white}0&\cellcolor{white}-$\lambda_1$&0\\
\hline
\hline
$\lambda_{12}$&$\lambda_{15}$&0  &-$\lambda_{9}$&-$\lambda_{11}$&$\lambda_{10}$&0&0&0&$\lambda_3$&-$\lambda_5$&$\lambda_4$&0&0&0&\cellcolor{white}-$\lambda_1$\\
\hline
$\lambda_{13} $&-$\lambda_{10}$&$\lambda_7$&0&0&-$\lambda_{15}$&$\lambda_{14}$&-$\lambda_2$&0&0&$\lambda_1$&0&0&0&-$\lambda_6$&\cellcolor{white}$\lambda_5$\\
\hline
$\lambda_{14}$ &-$\lambda_{11}$&$\lambda_8$&0&$\lambda_{15}$&0&-$\lambda_{13}$&0&-$\lambda_2$&0&0&$\lambda_1$&0&$\lambda_6$&0&\cellcolor{white}-$\lambda_4$\\
\hline\hline
\rowcolor{yellow}$\lambda_{15}$ & \cellcolor{white}-$\lambda_{12} $ &
 \cellcolor{white}$\lambda_{9}$ 
 &0 &\cellcolor{white}-$\lambda_{14}$ &\cellcolor{white}$\lambda_{13}$ &0&0&0&
\cellcolor{white}-$\lambda_2$ &0 &0&\cellcolor{white}$\lambda_1$
&\cellcolor{white}-$\lambda_5$ &\cellcolor{white}$\lambda_4$&0\\
\hline
\end{tabular}}
\vspace{0.5cm}
\caption{Commutator relations for $\mathfrak{su(4)}$.}
\end{sidewaystable}

\bigskip
\bigskip
The  block-diagonal form of  basis elements of subalgebra  $\alpha_X\,$   resulting under transposition $P_\pi$:

\begin{eqnarray}\label{eq:ChB1}
P_\pi \lambda_{3} P_\pi = \frac{i}{2}
\left(
\begin{array}{c|c}
{\sigma_3} & 0 \\
\hline
0 & -\sigma_3 \\
\end{array}
\right) , &\qquad&
P_\pi \lambda_{6} P_\pi = \frac{i}{2}
\left(
\begin{array}{c|c}
{\sigma_3} & 0 \\
\hline
0 & \sigma_3 \\
\end{array}
\right), \\
P_\pi \lambda_{7} P_\pi =  \frac{i}{2}
\left(
\begin{array}{c|c}
{\sigma_1} & 0 \\
\hline
0 & \sigma_1 \\
\end{array}
\right) , &\quad \quad &
P_\pi \lambda_{8} P_\pi = \frac{i}{2}
\left(
\begin{array}{c|c}
{\sigma_2} & 0 \\
\hline
0 & \sigma_2 \\
\end{array}
\right),  \\
P_\pi \lambda_{10} P_\pi = \frac{i}{2}
\left(
\begin{array}{c|c}
{\sigma_2} & 0 \\
\hline
0 & -\sigma_2 \\
\end{array}
\right),   &&
P_\pi \lambda_{11} P_\pi = \frac{i}{2}
\left(
\begin{array}{c|c}
{\sigma_1} & 0 \\
\hline
0 & -\sigma_1 \\
\end{array}
\right),\\
P_\pi \lambda_{15} P_\pi =\frac{i}{2}
\left(
\begin{array}{c|c}
I   & 0 \\
\hline
\medskip
0&  -I \\
\end{array}
\right) \,. &&
\label{eq:ChB2}
\end{eqnarray}

\newpage

\end{document}